\documentclass[12pt]{article}
\pdfoutput=1

\usepackage{graphicx}
\usepackage{psfrag}
\usepackage{enumerate}
\usepackage{amssymb}
\usepackage{jheppub}

\newcommand{\be}{\begin{equation}}
\newcommand{\ee}{\end{equation}}
\newcommand{\beq}{\begin{eqnarray}}
\newcommand{\eeq}{\end{eqnarray}}

\def\[{\left [}
\def\]{\right ]}
\def\({\left (}
\def\){\right )}

\def\r2{\sqrt{2}}

 \def\simleq{\; \raise0.3ex\hbox{$<$\kern-0.75em
      \raise-1.1ex\hbox{$\sim$}}\; }
   \def\simgeq{\; \raise0.3ex\hbox{$>$\kern-0.75em
      \raise-1.1ex\hbox{$\sim$}}\; }

\newcommand{\bbibitem}[1]{\bibitem{#1}\marginpar{#1}}

\newcommand{\figref}[1]{Fig.~\ref{#1}}

\def\Label#1{\label{#1}%
  \smash{\hbox to0pt{\raise1ex\hbox{\tiny[#1]}\hss}}}
\def\noLabels{\let\Label=\label}
\def\nobbibitem{\let\bbibitem=\bibitem}

\begin{document}

\noLabels
\nobbibitem

\DeclareGraphicsExtensions{.pdf,.png,.gif,.jpg,.eps}

\begin{titlepage}

\begin{center}
{\large \bf Observing the Multiverse with Cosmic Wakes}

\vspace{6mm}

{Matthew Kleban$^1$, Thomas S. Levi$^2$, Kris Sigurdson$^2$}

\vspace{6mm}
{\it $^1$Center for Cosmology and Particle Physics\\
New York University \\
New York, NY 10003, USA}

 \vspace{6mm}
{\it $^2$Department of Physics and Astronomy\\
 University of British Columbia\\
  Vancouver, BC V6T 1Z1, Canada }

\end{center}
%------------------------------------------------------------------------------

%\setcounter{footnote}{0}
%%%%%%%%%%%%%%%%%%%%%%%%%%%%%%%%%%%%%%%%%%%%%%%%%%%%%%%%%%%%%%%%%%%%%%%%%%%%%%%%%%%%%%%
\begin{abstract}
\noindent

Current theories of the origin of the Universe, including string theory, predict the existence of a multiverse containing many bubble universes.  These bubble universes will generically collide, and collisions with ours produce \emph{cosmic wakes} that enter our Hubble volume, appear as unusually symmetric disks in the cosmic microwave background (CMB) \cite{wwc2} and disturb large scale structure (LSS). There is preliminary observational evidence consistent with one or more of these disturbances on our sky \cite{Feeney:2010jj}. However, other sources can produce similar features in the CMB temperature map and so  additional signals are needed to verify their extra-universal origin. Here we find, for the first time, the detailed three-dimensional shape and CMB temperature and polarization signals of  the cosmic wake of a bubble collision in the early universe consistent with current observations \cite{Komatsu:2010fb}. The predicted polarization pattern has distinctive features that when correlated with the corresponding temperature pattern are a unique and striking signal of a bubble collision. These features  represent the first verifiable prediction of the multiverse paradigm and might be detected by current experiments such as Planck \cite{planck} and future CMB polarization missions. A detection of a bubble collision would  confirm the existence of the Multiverse, provide compelling evidence for the string theory landscape, and sharpen our picture of the Universe and its origins.

\end{abstract}
%%%%%%%%%%%%%%%%%%%%%%%%%%%%%%%%%%%%%%%%%%%%%%%%%%%%%%%%%%%%%%%%%%%%%%%%%%%%%%%%%%%%%%%%%
%\vspace{0.5in}
\end{titlepage}

%\renewcommand{\baselinestretch}{1.05}  %Line spacing
%%%%%%%%%%%%%%%%%%%%%%%%%%%%%%%%%%%%%%%%%%%%%%%%%%%%%%%%%%%%%%%%%%%%%%%%%%%%%%%%
%%%%%%%%%%%%%%%%%%%%%%%%%%%%%%%%%%%%%%%%%%%%%%%%%%%%%%%%%%%%%%%%%%%%%%%%%%%%%%%%%%%%%%%%%%%
%\tableofcontents 

\section{Introduction}

\begin{figure}
\hspace{-0.1 in}\includegraphics[width=1.0\textwidth]{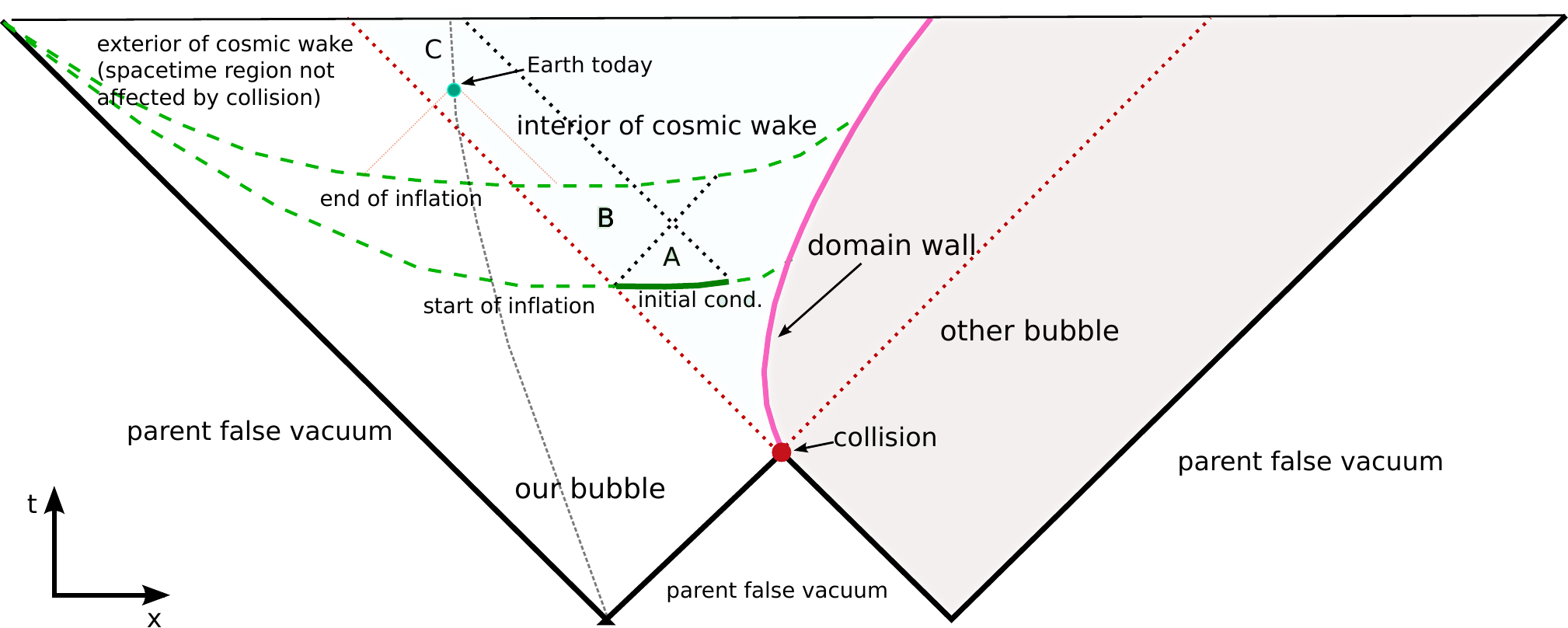}
\caption{\label{st} Spacetime conformal diagram of a cosmic bubble collision.  Each point represents a two-dimensional hyperbolic slice of space with a radius of curvature that depends on position within the diagram.  The physics in regions A, B, and C is determined by the conditions on the surface labelled ``initial cond.'' inside the cosmic wake and by the conditions outside the cosmic wake.   To a good approximation, the physics in regions B and C depends only on the left-moving part of the inflaton perturbation.}
\end{figure}

The possibility of observing cosmic bubble collisions has recently received a considerable amount of attention  (see, e.g., \cite{Kleban:2011pg} for a recent review).  Such collisions are a generic prediction of multiple-vacua models like the string theory landscape, and an observation of one would fundamentally alter our understanding of the cosmos at large scales.  In these models the observable part of the Universe is contained within a bubble that formed as a result of a first-order phase transition from a parent false vacuum.  Collisions with other such bubbles produces a special wave, that we term a {\it cosmic wake}, that propagates into our bubble and affects the spacetime region to the causal future of the collision (see \figref{st}).  

To a remarkable extent these dramatic events can be analyzed analytically and in a model-independent fashion.
In this work we determine the precise pattern of temperature and polarization of Cosmic Microwave Background (CMB) photons induced by a cosmic wake using the full solution to the cosmological Einstein-Boltzmann equations. The results, while consistent with a previous analytic approximation \cite{Czech:2010rg}, are remarkable: collisions can produce a unique and highly characteristic polarization signal---a ``double peak'' in the magnitude of the polarization as a function of angle.  This double peak, and more generally the large-scale azimuthally symmetric polarization and temperature pattern produced by cosmic bubble collisions, serves as a true smoking gun for their detection.  Moreover, in an interesting regime of parameters this vital corroborating signal in CMB polarization can be as easy or potentially even easier to detect than the signal in CMB temperature.  

Our results also determine the evolution of density and velocity perturbations after decoupling, which opens the door for future work  quantifying the effect of the cosmic wake on large scale structures.

\section{Background}

While the detailed physics of cosmic bubble collisions depends on high-energy microphysics, to a surprising extent the physics relevant to the late-time cosmic wake that affects observables such as the CMB radiation and Large Scale Structure (LSS) appears to depend on only a few simple parameters.  
This occurs because the collision between two bubbles preserves much of the symmetry inherited from bubble nucleation events, and the inflationary period in our Universe 
subsequent to any collision rapidly erases all but a few characteristic features.

In a broad class of microphysical models there are only four parameters relevant to the effects of a collision on the observable Universe at the end of inflation.  Of these, one probes the underlying microphysics, while the others reflect the initial conditions (the position and time of the nucleation of the colliding bubble relative to ours).  These four parameters determine the direction, distance to, and strength of the collision-induced cosmic wake in our Hubble volume, and therefore the location, size, and intensity of the affected disc in the CMB sky.  

Because the primordial plasma supports acoustic waves, the cosmic wake can propagate into our Hubble volume.
The most dramatic effect is the appearance of two sharp peaks in the magnitude of E-mode polarization as a function of radius:  there is a double concentric ring of strongly polarized light outlining the affected disk in the CMB sky.  These peaks are a position-space manifestation of the acoustic physics responsible for the more familiar peaks observed in the angular temperature power spectrum of the CMB.

\subsection{Scalar Perturbations after Inflation}

We first calculate the expected curvature perturbations resulting from cosmic bubble collisions that persist after a period of slow-roll inflation.  These perturbations subsequently evolve into the cosmic wake.

The spacetime metric and scalar-field configuration 
describing the collision of two Coleman-de Luccia thin-wall bubbles expanding in a parent false vacuum preserves an $SO(2,1)$ group of isometries \cite{Hawking:1982ga, ggv, Freivogel:2007fx, wwc}.  
In suitable coordinates the collision takes place at one instant of time everywhere along a hyperbolic surface ${\mathbb H}_2$ in space, and the effects of the collision causally influence the spacetime region to the future of the surface at that instant.

During inflation the background metric describes, up to slow-roll corrections, an approximate de Sitter spacetime with Hubble constant $H_I$ and may be written in the form $ds^2 = -(1+H_I^2t^2)^{-1} dt^2 +(1+H_I^2t^2) dx^2 + t^2 d{\mathbb H}_2^2$, where $d{\mathbb H}_2^2 = dr^2 + \sinh^2 r d \varphi^2$ is the metric on the surface ${\mathbb H}_2$.  Since the collision preserves the $SO(2,1)$ symmetry its effects are uniform on the ${\mathbb H}_2$.  

However, after $N$ $e$-folds of inflation the coordinate  $H_I t \propto e^N$ becomes exponentially big and the hyperbolic curvature small.  Current constraints indicate that $R_0 \simgeq 10c/H_0$, or equivalently $\sqrt{\Omega_k} = (R_0H_0/c)^{-1}\simleq 0.1$,  where $R_0$ is the radius of curvature of the Universe today and $\Omega_k$ is the effective curvature density \cite{Komatsu:2010fb},
so it is a good approximation to ignore the negative spatial curvature and treat the hyperbolic surface ${\mathbb H}_2$ as planar (which we do for the remainder of this work).  We also  approximate the inflationary spacetime as de Sitter space.
Curvature and higher-order slow-roll effects induce small corrections to the result calculated here.  Finally, we use the thin-wall approximation for the collision bubble---that is, we treat the region affected by the collision as having a sharp boundary.  Taking into account finite thickness will have the effect of smearing out very small scale features in the signal, and sufficient smearing would replace the double peak with a single peak.  The extent to which this happens is a model-dependent question and could provide further details about the microphysics---specifically, if a distinct double peak is observed it indicates that the ratio of the Hubble scale during inflation $H_I$ to the Hubble scale of the parent false vacuum $H_F$ satisfies $(H_I/H_F)^2 \simleq 10^{-3}$  \cite{gob}.

In the planar approximation and including linear perturbations in conformal Newtonian gauge, the metric takes the form
\be
ds^2 = (H_I \tau)^{-2} \left[ -(1+ 2 \Phi) d\tau^2 + (1-2 \Psi) (dx^2 + dy^2 +dz^2) \right],
\ee
where $H_I$ is the Hubble parameter during inflation, $\tau < 0$ is the conformal time, the spatial coordinates are chosen so that the solar system's comoving location is $\vec x = 0$, and $\Phi$ and $\Psi$ are the Newtonian potentials (which are equal during inflation). 
In the approximation that the collision surface is planar, the region of spacetime affected by the collision satisfies $x > -\tau + x_c$ for some constant $x_c$ that depends on when the collision occurred.   We focus on collisions with a region of causal influence that intersects the visible part of the surface of last scattering because these have the best potential for detection \cite{wwc2, bubmeas, Czech:2010rg}; since the Earth is at $x=0$ and $H_I |\tau| \ll 1$ at the end of inflation, we take $H_I |x_c|  \ll 1$.  

Given that slow-roll inflation took place,  pre-inflationary inhomogeneities became small at some time early in inflation.  From that time on we may evolve perturbations using standard linear cosmological perturbation theory \cite{mfb}.  To lowest order in the slow-roll parameters, gauge-invariant perturbations to the inflaton field $\phi = \phi_{0}+\delta \phi$ obey the equation of motion of a free scalar in de Sitter space \cite{mfb}
\be \label{MFB}
\partial_{\tau}^{2 } \delta \phi + 2 {\cal H} \partial_{\tau} \delta \phi -\nabla^2 \delta \phi + {\cal O}(\epsilon, \eta)=0,
\ee
where ${\cal H}(\tau)=-1/\tau + {\cal O}(\epsilon, \eta)$ is the conformal Hubble constant, and $\epsilon, \eta$ are the standard slow-roll parameters.    We neglect terms of ${\cal O}(\epsilon, \eta)$ and work at zeroth order in the slow roll expansion.

Because the collision perturbation is constant on the nearly-planar surface the general solution to
 \eqref{MFB}  can be written in closed form \cite{wwc2}
\be \label{infsol}
\delta \phi(\tau,x)= \tilde g(\tau+x) - \tau \tilde g'(\tau + x)+ \tilde f(\tau-x) - {\tau}\tilde f'(\tau-x),
\ee
where $\tilde f$ and  $\tilde g$ are arbitrary functions of one variable (and $\tilde f'$ and $\tilde g'$ their derivatives with respect to their argument). 
To proceed further we need the initial perturbation $\delta \phi$ produced by the bubble collision event.  
Without a specific model for the underlying microphysics this cannot be computed but, to a good approximation, the post-inflation predictions of any such model can be characterized by four parameters.

Consider the region near the edge of the collision lightcone $x =- \tau_i + x_c$ at some time $\tau_i \sim -H_I^{-1}$ near the beginning of inflation when linear perturbation theory is valid (labelled ``initial cond.'' in \figref{st}).  Since the perturbation $\delta \phi$ is non-zero only for $x >- \tau_i + x_c$, from \eqref{infsol} we must have $\tilde g(\tau + x-x_c) = g(\tau + x-x_c) \Theta(\tau+x-x_c)$ and $ \tilde f(\tau-x) = f(-\tau+x-x_c+2 \tau_i)\Theta(-\tau+x-x_c+2 \tau_i)$, where $\Theta$ is the Heaviside step function and $f,g$ are functions of one variable.  
The $\tilde f$ terms in \eqref{infsol} are rightmoving excitations.  By the end of inflation $H_I |\tau_e| \ll  1$, they are non-zero only for $x-x_c > -2 \tau_i \sim 2 H_I^{-1}$ but,  as explained above, we are only interested in $|x - x_c| \ll H_I^{-1}$ and these terms are not relevant for cosmological observables ({\it c.f.} \figref{st}, specifically regions B and C).

At the end of inflation, the inflaton perturbation is \cite{gob}
\be \label{geq}
\delta \phi(\tau_e, x) \approx g(x-x_c)\Theta(x-x_c) =  M \sum_{n=0}^\infty \left\{ {\alpha_n} H_I^n \( x-  x_c\)^n \right\} \Theta(x-x_c),
\ee
where we have dropped the term in \eqref{infsol} proportional to $\tau_e  \ll H_I^{-1}$, and expanded $g$  in terms of dimensionless coefficients $\alpha_n$ and a constant $M$ with dimensions of mass.

The model-dependent initial conditions $\delta \phi(\tau_i, x)$ and $\delta \dot \phi ( \tau_i, x)$ determine the coefficients $\alpha_n$. 
As is clear from \eqref{infsol}, any regular perturbation $\delta \phi$ gives $\alpha_0=0$.  For an ${\cal O}(1)$ perturbation at an early time $\tau_i \sim H_I^{-1}$ one generically expects the dimensionless coefficients $ \alpha_n \simleq 1$ for $n>0$. 
But as we are only interested in the region $H_I |x-x_c|  \ll 1$, absent fine-tuning of the $\alpha_n$, the higher terms in \eqref{geq} will be negligible.  Thus, by the end of inflation, the leading perturbation reduces to just  the linear term \cite{wwc2, gob}
\be \label{phieq}
\delta \phi = M \alpha_1 (x-x_c) \Theta(x-x_c).
\ee
This expression contains two parameters ($M \alpha_1$ and $x_c$).  We have chosen coordinates so that the $x$ direction is the axis of symmetry of the collision; {\it i.e.} there are two angles parametrizing the direction towards center of the collision.  As mentioned above, these four parameters fully characterize the effects on the CMB of a generic collision.

\subsection{Curvature Perturbations after Inflation}

The inflaton perturbation $\delta \phi$ determines the conformal Newtonian gauge gravitational potential perturbation $\Phi$. Using the Einstein and scalar equations we can show \eqref{phieq} leads to a potential perturbation
\be \label{phipert}
\Phi = - {V' \over 2} g(\tau+x)\Theta(\tau+x) \approx -{V' \over 2} M \alpha_1 x \Theta(x),
\ee
where $V'=\partial V(\phi)/\partial \phi$ is the slope of the inflationary potential.

The initial conditions for cosmology after inflation are best expressed in terms of the comoving curvature perturbation $\zeta$ that is conserved on super-Horizon scales \cite{mfb,Lyth:1984gv}. 
We find
\be \label{curv-pert}
\zeta = {2 \over 3} {{\cal H}^{-1} \partial \Phi/ \partial \tau + \Phi \over 1+w} + \Phi =  {- M \alpha_1 V' \over 3}\left( {  x  \over 1+w} + \tau + x \right) \Theta(\tau+x) \approx  \lambda x\Theta(x),
\ee
where $w\simeq -1$ is the equation-of-state during inflation and $\lambda = -\alpha_1 M V'(2+w)/(3+3w) \sim \alpha_1 M V/V'$ is a constant that sets the amplitude of late-time effects. The final, approximate equality is valid late in the inflationary epoch.

\section{Results}

Using the superhorizon curvature perturbation $\zeta_i$ near the end of inflation and a set of {\it linear transfer functions} $\tilde \Delta_X $ the perturbed distribution of an observable $X$ at later times is
\beq \label{transfer defined}
\Delta_X ({\bf x}, \hat{n}, \tau) = \int d^3 k \, e^{i {\bf k} \cdot {\bf x} } \tilde{\Delta}_X (k=| {\bf k} | , \hat{n}, \tau) \zeta_i ( {\bf k} ) ,
\eeq
where $\Delta_X ({\bf x}, \hat{n}, \tau)$ is the local value of an observable $X$ (e.g. $\delta T / T$ for the photon distribution) 
at position ${\bf x}$, and time $\tau$ in the direction  $\hat{n}$ on the sky, and ${\bf k}$ is the comoving wave-vector. 
Angular moments of  $\Delta_X ({\bf x}, \hat{n}, \tau)$ determine, for example, the density and velocity as a function of position, while  $\Delta_X (0, \hat{n}, \tau)$ is the angular anisotropy at the location of the Earth.

The transfer functions encaspulate the full evolution of the coupled multi-component fluid and gravitational system and to linear order depend only on the background cosmology---hence they are independent of the collision perturbation. For the cosmological background we choose best-fit values from the 7-year data release of WMAP \cite{Komatsu:2010fb}.
We compute the transfer functions using a customized version of CAMB \cite{Lewis:1999bs}, and our own code to numerically evaluate the Fourier transform in \eqref{transfer defined} and extract observables.

\subsection{Temperature} 

In the Sachs-Wolfe  approximation \cite{sw}, the CMB temperature anisotropy is determined by the curvature perturbation at decoupling: $\delta T (\theta, \phi)/T\propto \zeta(\vec x, \tau_{dc})$, where $\vec x^{\, 2}=D_{dc}^2$ is the distance to the last scattering surface and the angular coordinates are chosen so that $x=D_{dc} \cos \theta$.  In this approximation the collision perturbation \eqref{curv-pert} gives rise to a very simple temperature perturbation: $\delta T/T \propto \Theta(\theta_c-\theta) \( \cos \theta - \cos \theta_c \)$, where $\theta_c=\cos^{-1} x_c/D_{dc}$  is the angular radius of the spot \cite{wwc2,Czech:2010rg}.

However, the perturbation \eqref{curv-pert} is constant only on super-horizon scales after the end of inflation.  To take into account the full evolution between reheating and decoupling, we instead use \eqref{transfer defined}.
\begin{figure}
\hspace{-0.1 in}\includegraphics[width=1.0\textwidth]{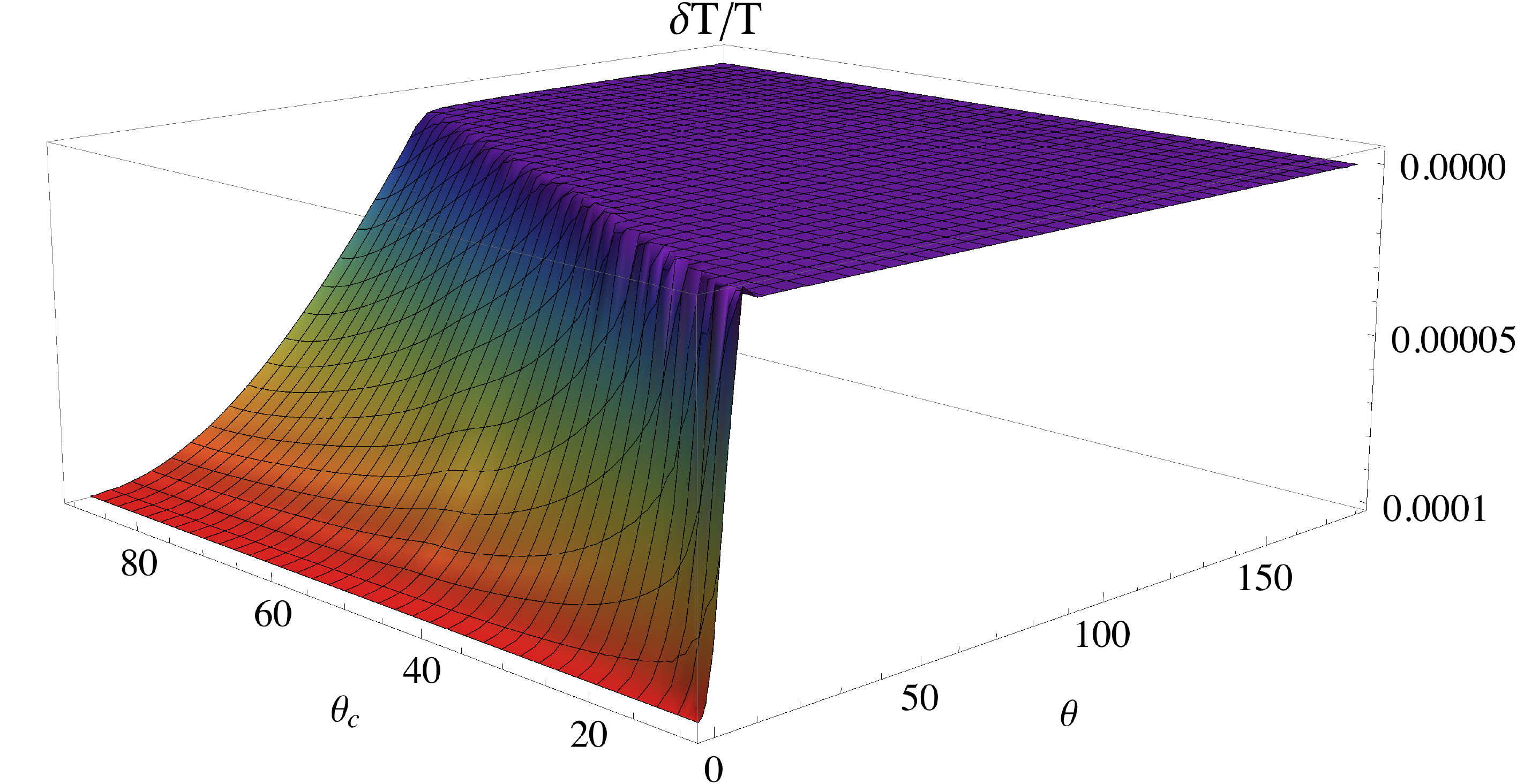}
\caption{\label{fig-temp} Plot of $\delta T / T$ for all possible spot angular radii from $\theta_{c}=5^\circ$ to $90^\circ$. The spots are normalized so that the $|\delta T / T| = 10^{-4}$ at the center of the spot.   }
\end{figure}
The results are presented in \figref{fig-temp}, with $\theta=0$ pointing along the $x$-axis (towards the collision) and $|\delta T / T |$ normalized to $10^{-4}$ at the center of the spot. We present a range of $x_c$ on the figure: the $y$-axis is the approximate angular radius of the spot $\theta_{c} \approx \cos^{-1} x_c/D_{dc}$, and  the $x$-axis is the angle from the spot center.

For all $x_c$, we find the temperature perturbation has the largest magnitude at the center of the spot $\theta=0$, decreases linearly in $\cos \theta$, and then smoothly transitions to a constant at $\theta=\theta_c$ (with no edge or discontinuity).  The width in $\theta$ of the transition region is roughly equal to the angular separation between the two peaks in polarization discussed below.  These results are fully consistent with the approximate analytic results of \cite{wwc2,Czech:2010rg}.  Whether the spot is hot or cold encodes microphysics of the inflaton and the collision bubble, see \cite{wwc2}.

\subsection{Polarization} 

\begin{figure}
\begin{center}$
\begin{array}{cc}
\hspace{-.5in}\includegraphics[width=0.38\textwidth]{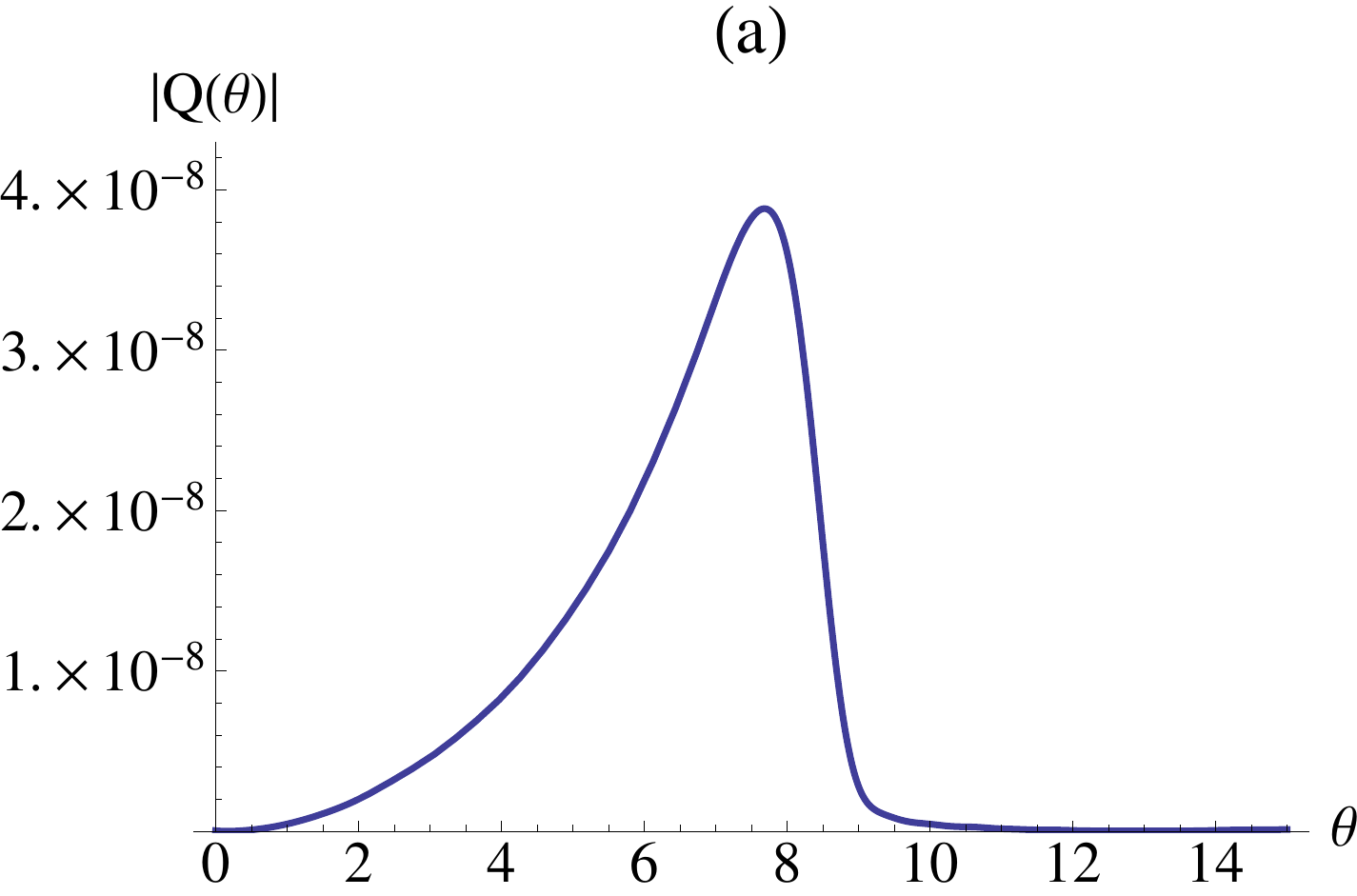}
\includegraphics[width=0.38\textwidth]{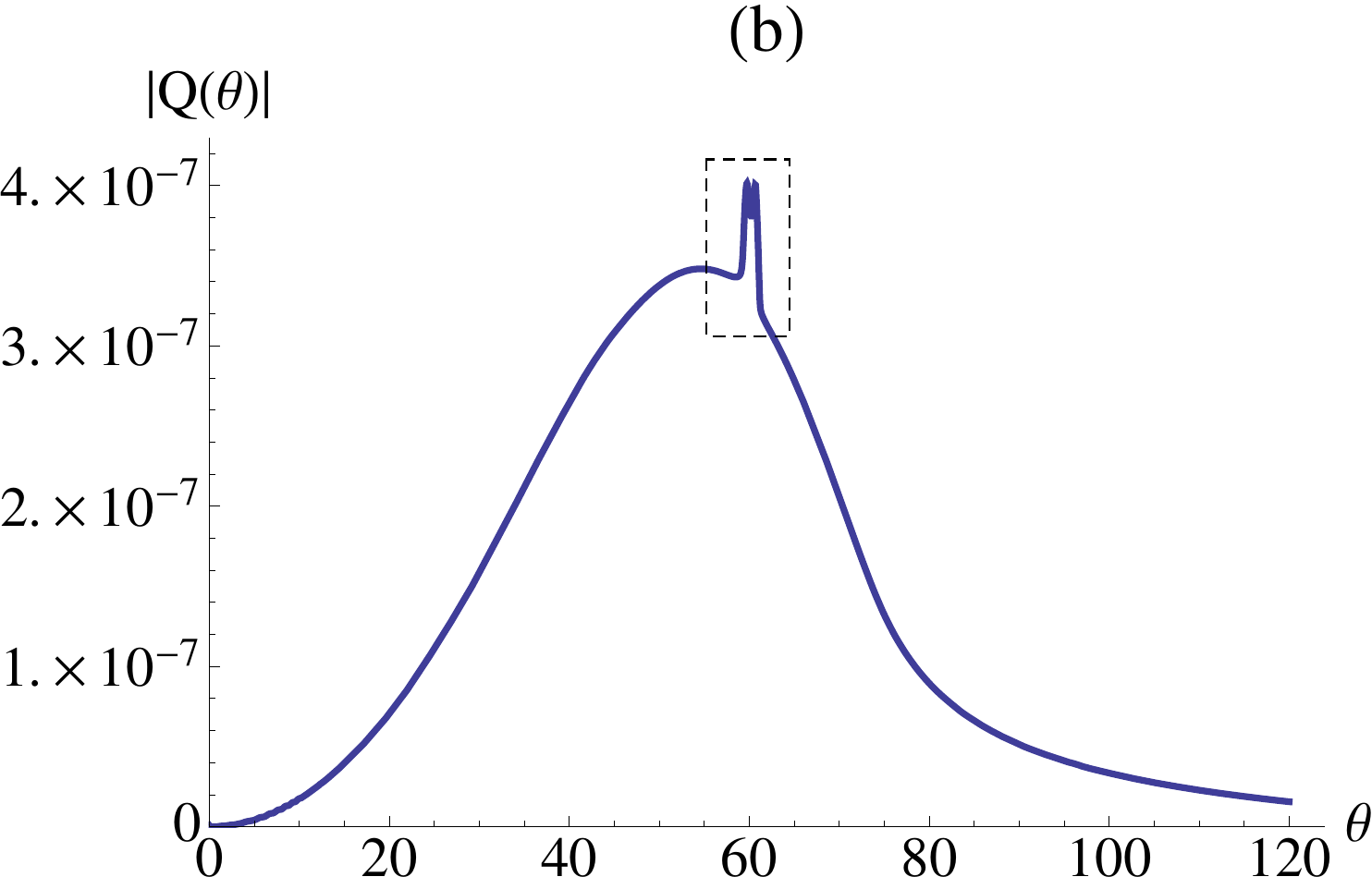} 
\includegraphics[width=0.38\textwidth]{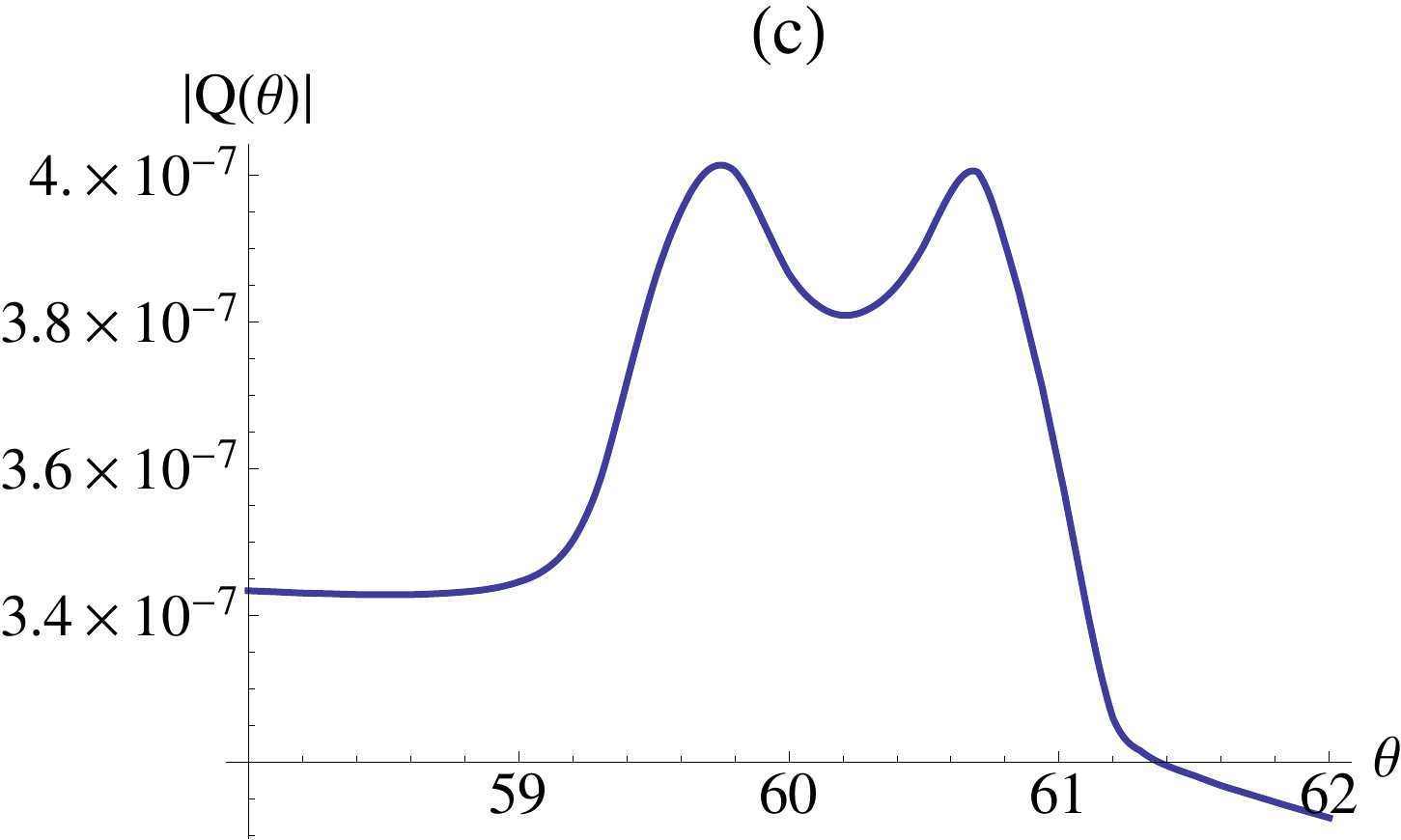}
\end{array}$
\end{center}
\caption{\label{fig-polcross} Polarization $|Q(\theta)|$ for (a): a spot with  angular radius $\theta_c \approx 10^\circ$, and (b): a spot with angular radius $\theta_c \approx 60^\circ$. Note the double peak in (b), with the indicated region magnified in (c). The spots are normalized so that  $|\delta T/T| = 10^{-4}$ at the center. }
\end{figure}

The CMB is linearly polarized due to Thomson scattering of CMB photons off free electrons \cite{Bond:1984fp,Zaldarriaga:1996xe}. This scattering occurs primarily at redshifts around decoupling ($z_{dc} \sim 1100$) and reionization ($z_{re} \sim 10)$. Scattering produces a net polarization proportional to the quadrupole temperature anisotropy of the radiation incident on the electron. Scattering at decoupling depends on well understood atomic physics \cite{Seager:1999bc}, but there is significant uncertainty in the reionization history of the Universe. However, our key results are not very sensitive to different reionization models \cite{Czech:2010rg}, and we choose a simple model of single reionization as our fiducial case \cite{Komatsu:2010fb}.

The polarization of a transverse electromagnetic wave with intensity tensor $I_{ij}$ can be characterized by the Stokes parameters $Q$ and $U$:
\beq
Q = (I_{11} - I_{22})/4, \quad U = I_{12}/2.
\eeq
Choosing coordinates so that the collision spot is centered on $\theta=0$ guarantees that the Stokes parameter $U=0$ ({\it i.e.} the polarization is purely $E$-mode, as expected for a scalar perturbation). The polarization pattern is radial (azimuthal) for a cold (hot) spot with $Q > (<) \,0$. For a discussion of CMB polarization in general see \cite{Zaldarriaga:2003bb}.

\begin{figure}
\hspace{.2 in}  \includegraphics[width=0.78\textwidth]{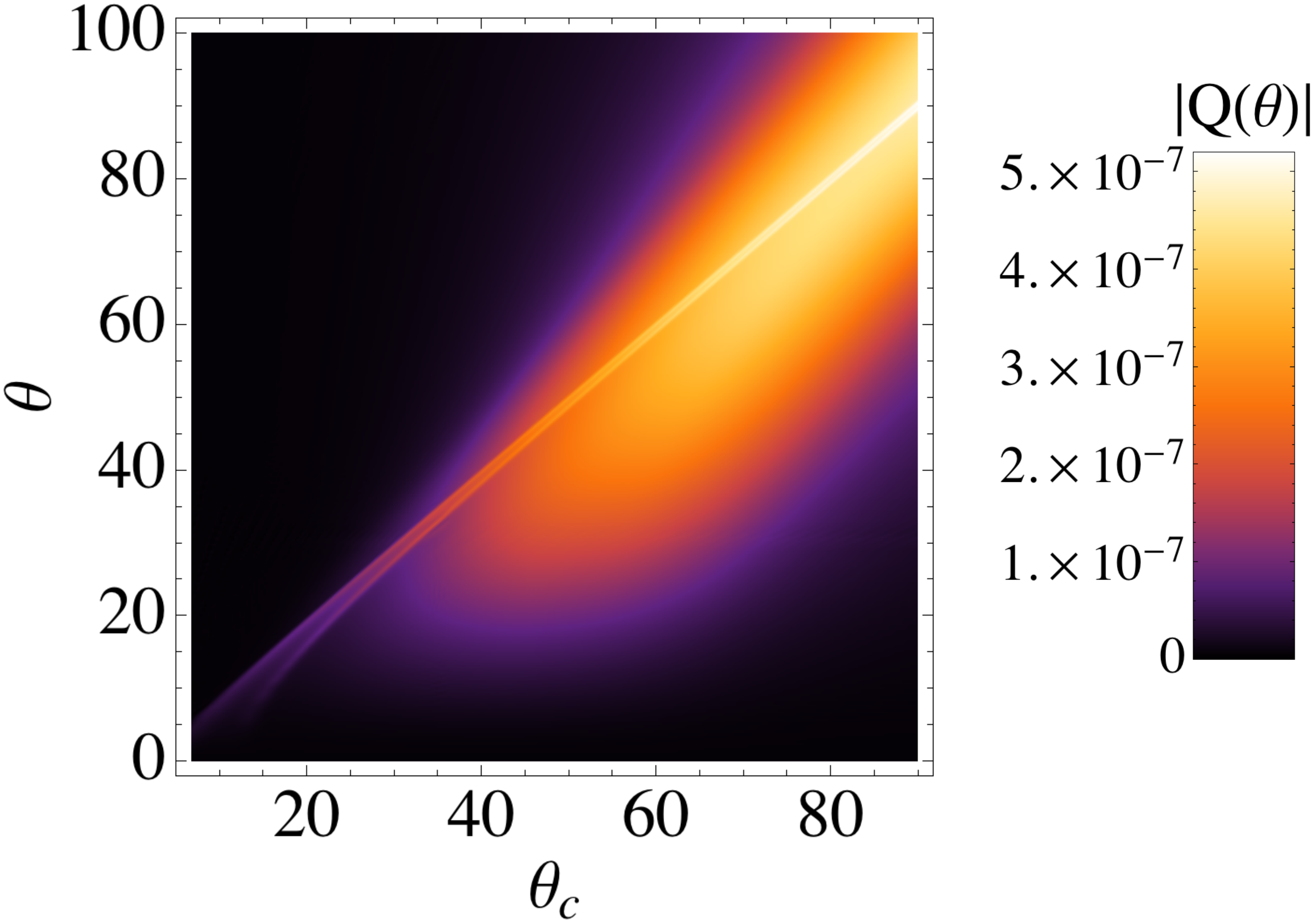}
 \caption{\label{fig-poldensity}  Density plot of the polarization $|Q(\theta)|$ for spot radii from  $\theta_{c} = 5^\circ$ to $90^\circ$. The plot is normalized such that each spot has a temperature contrast $|\delta T/T| = 10^{-4}$ in the center.}
\end{figure}

Thanks to the azimuthal symmetry, we can parametrize the CMB polarization due to a single collision by $Q=Q(\theta)$ alone.
However $Q(\theta)$ differs qualitatively for small ($\theta_c \simleq 12^\circ$) and large spots. We display two cases in \figref{fig-polcross}; $\theta_c \approx 10^\circ$ and  $\theta_c \approx 60^\circ$. The large spot has a distinctive ``double peak" structure; two rings of sharply enhanced polarization concentric with the edge $\theta=\theta_c$ of the affected disk in the temperature map.  
This structure is characteristic of bubble collisions and appears for any spot with $\theta_{c} \simgeq 12^\circ$.  

In \figref{fig-poldensity} we display the results for all collisions with $5^\circ< \theta_{c} <90^\circ$. The $x$-axis is $\theta_{c}$, the $y$-axis is $|Q(\theta)|$.

\subsubsection{Why a double peak?} The double peak in \figref{fig-polcross} and \figref{fig-poldensity} is a very striking feature, but the physics that gives rise to it is simple. 
At the end of inflation, the perturbation \eqref{phipert} in the Newtonian potential is $\Phi  \sim x \, \Theta(x)$. Between the end of inflation and decoupling, a linear-in-$x$ perturbation (such as $\Phi = C x$)  remains so, because the evolution equations for cosmological perturbations are all second order in spatial derivatives.  By contrast, the ``kink" at $x=0$ spreads at the sound-speed $c_s \lesssim c/\sqrt{3}$ in the plasma, and evolves into a smooth profile confined within the  sound horizon of $x=0$ until matter-radiation equality.  After equality the speed of sound is effectively zero and there is no further spreading. However at the edge of the affected region at any time, the first derivative of the perturbation is still close to discontinuous. 

\begin{figure}
\hspace{1.4in} \vspace{4 in} \includegraphics[width=0.5\textwidth]{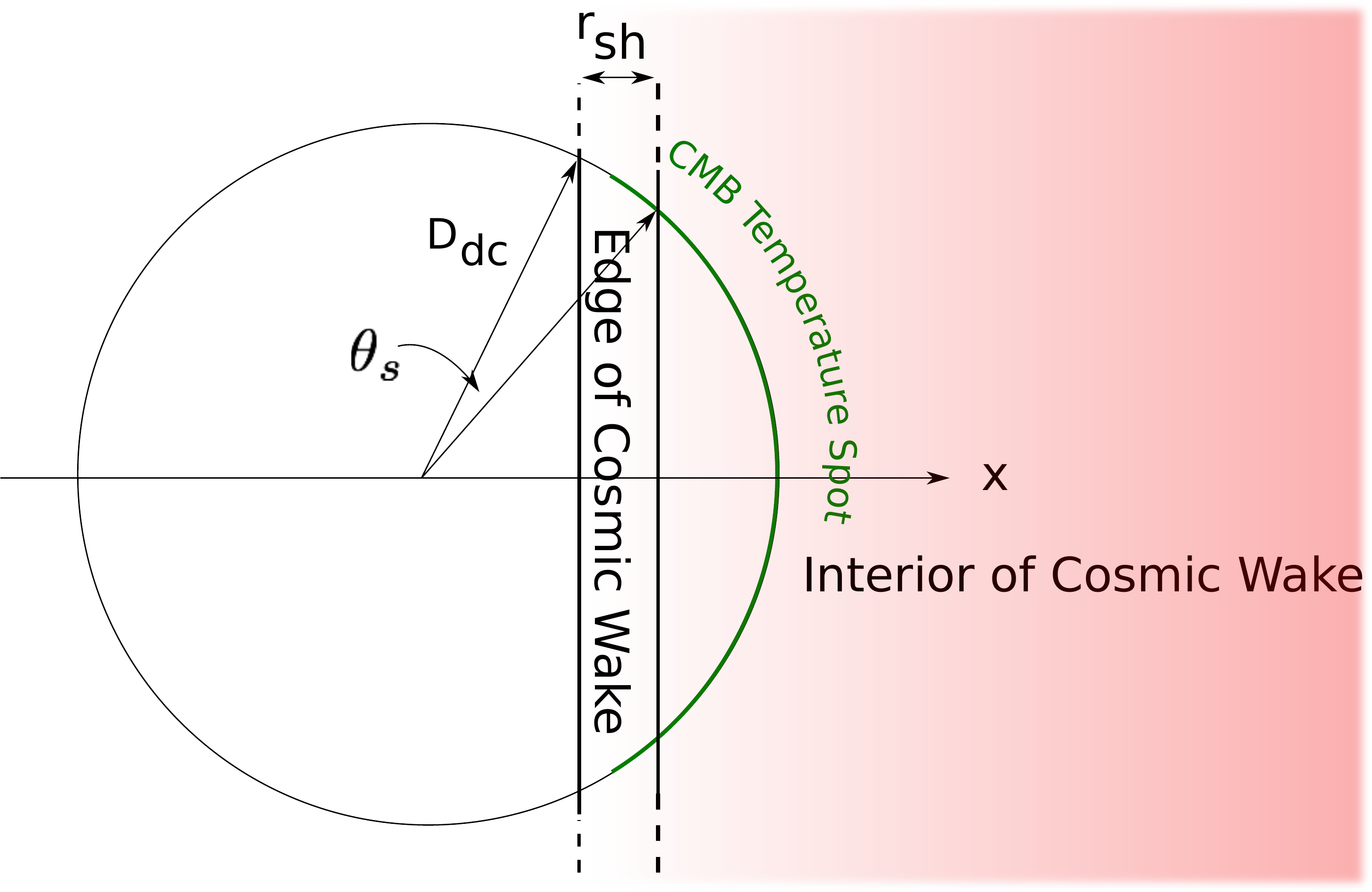}
 \vspace{-4 in}  \caption{\label{dplc}  The last scattering surface and the cosmic wake, showing the region of the CMB temperature sky affected by the collision and illustrating the geometric origin of the double peak separation angle.}
\end{figure}

An electron with a last scattering sphere that intersects one of the edges of this region will scatter a relatively large amount of polarized light, because the quadrupole moment is sensitive to the second derivative of the temperature distribution.  The two edges of the region at decoupling should be separated by twice the sound horizon $2r_{sh}\sim 306$ Mpc, or $r_{sh}/D_{dc}\approx .0109$ \cite{Komatsu:2010fb} (see \figref{dplc}).
 Therefore the angular separation between the two edges should be $\theta_s \approx \cos^{-1}\(x_c-r_{sh} \)/D_{dc}-\cos^{-1}\(x_c+r_{sh} \)/D_{dc}$, which agrees well with our numerical results.  For $\theta_c \approx 90^\circ$ this width is $\theta_s \approx 2 \theta_{sh} = 1.3^{\circ}$.  For smaller disks, $\theta_s$ is larger, and for $\theta_c \simleq 12^{\circ}$ the inner peak disappears entirely.

\subsection{Density}

After matter-radiation equality, the sound speed of perturbations is approximately zero and the edge of the cosmic wake should remain at fixed $x$.  Due to the growth of perturbations during matter domination, the amplitude of the perturbation at the edge of the wake today should be amplified by roughly a factor of $z_{dc}\sim 1100$ from its value at decoupling.  Since a Newtonian potential of the form $\Phi = C x \theta(x)$ corresponds to a $\delta$-function density perturbation at $x=0$, one expects a planar sheet of over- or under-density (for hot or cold spots in the CMB, respectively) centered at $x \approx D_{dc} \cos \theta_c$, with a thickness set roughly by the sound horizon $r_{sh}$ at equality.

\begin{figure}
\hspace{0.3 in}\includegraphics[width=0.9\textwidth]{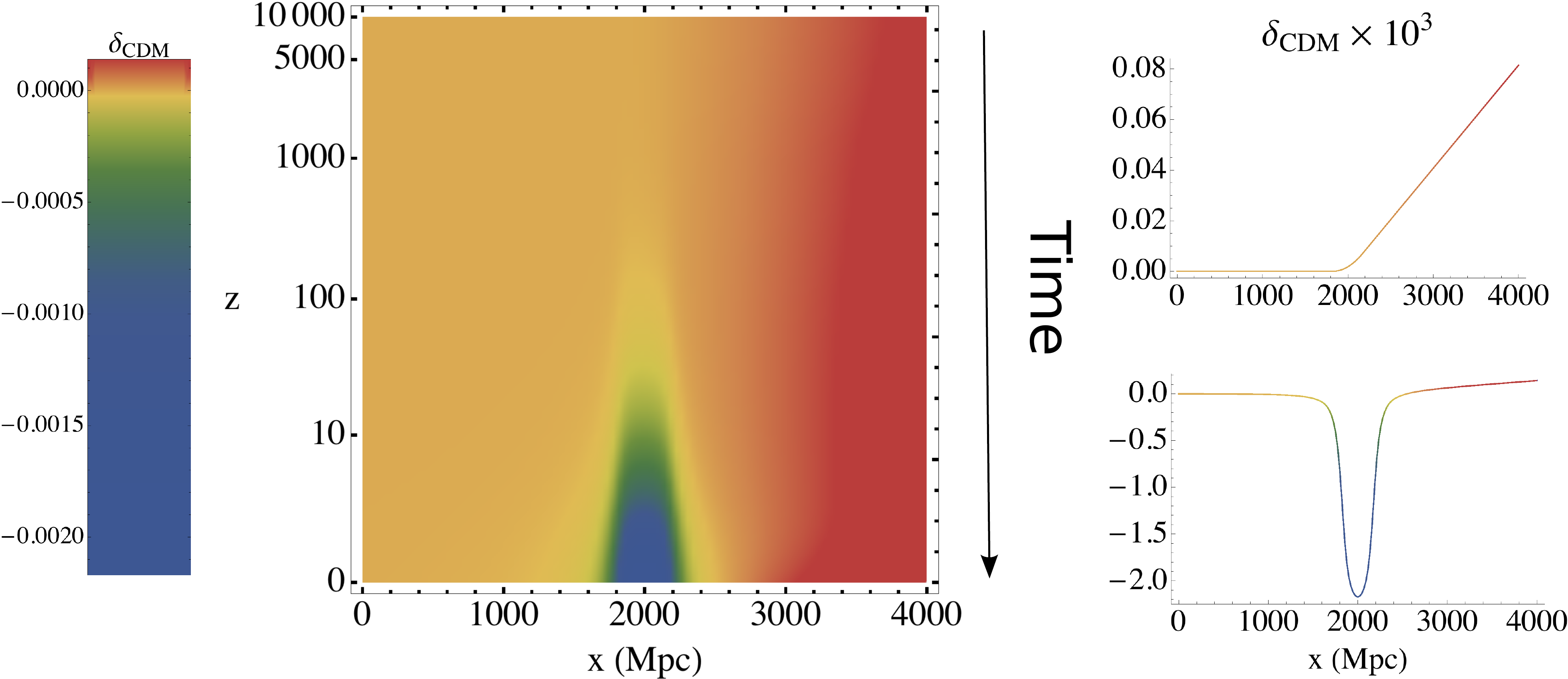}
\caption{\label{fig-cdm} Evolution of the cold dark matter density contrast  $\delta_{\rm cdm}$ in conformal Newtonian gauge,  from redshift $z=10,000$ to the present. The left panel shows the continuous evolution, while the right panel shows two snapshots at $z=10,000$ (top) and $z=0$ (bottom). The cosmic wake is normalized so that  $\delta T/T = -10^{-4}$  (a cold spot) at the center of an affected disk in the CMB with angular radius $\theta_c \approx 80^\circ$. }
\end{figure}

These expectations are confirmed numerically in \figref{fig-cdm}, where we display the cold dark matter density contrast in conformal Newtonian gauge (the baryon density looks similar).  The presence of such a mass sheet could lead to  signatures in large scale structure or lensing surveys which---combined with CMB data---could further corroborate the discovery of a bubble collision.  We  leave this investigation to future work \cite{lam}.

\section{Detectability}

In this section we estimate the degree of detectability of the polarization signal for spots of varying size and brightness and for a selection of current and future experiments.
Both the temperature and polarization signals are circularly symmetric. We orient our coordinates such that the center of the spot is at the pole $\theta = 0$; a  coordinate rotation can be used to center the spot anywhere else. 

To get an estimate of the detectability threshold we will fit to a one parameter model (which we label $A$) for the amplitude of the temperature spot at its center. We analyze the fit for temperature (T) alone, E-mode polarization (E) alone, and E-mode polarization combined with information from the temperature map and cross-correlation. We work in $l$-space where the noise covariance matrix is diagonal.   The circular symmetry allows us to integrate over the azimuthal angle $\varphi$, which in $l$-space means the $a_{l,m} ^{T,E}$ multipoles do not contribute for $m \neq 0$. 

For each estimate we generate a full sky of CMB fluctuations in T and E consistent with WMAP-7 concordance parameters \cite{Komatsu:2010fb} as well as a noise realization for each detector up to $l_{max} = 2000$. We then add a collision spot to this and compute the relative likelihood function (which amounts to a simple $\chi^2$ in this one parameter model) to determine if the spot is detectable.  We normalize such that for a given spot a null detection is $A=0$ (i.e. consistent with Gaussian fluctuations) and a perfect detection is $A=1$. As a sample we look at the projected sensitivities for three experiments: a current satellite mission (Planck \cite{planck}), a planned balloon mission (SPIDER \cite{spider2}) and a future satellite mission (CMBPol (EPIC-2m) \cite{Baumann:2008aq}).  We use the same predicted sensitivies, beam-widths, weight per solid angle ($w$), and observing time as \cite{Ma:2010yb}. For brevity we analyze just these three, but any experiment which measures the temperature and/or polarization over a significant fraction of the sky---such as CLASS, EBEX \cite{Oxley:2005dg}, PIXIE \cite{Kogut:2011xw}, ACTPol \cite{Niemack:2010wz}, and SPTpol \cite{Carlstrom:2009um}---has the potential to detect a cosmic wake.  

Since we are fitting for the $a_{lm}$ themselves rather than the power spectra $C_l$, a cut sky introduces straightforward but bothersome complications (the spherical harmonics are no longer orthonormal). To get a simple idealized estimate we use a full sky for each. For the satellite experiments this amounts to ignoring foreground effects in our estimates. For partial sky-coverage balloon missions we are assuming that the Gaussian fluctuations will be similar in the region of the sky not measured. A full treatment of these effects would likely decrease the quoted sensitivities by a modest factor.

\begin{table}
\caption{Projected detector parameters}
\centering
\footnotesize
\begin{tabular} {| c | c | c  | c |   }
\hline \hline

Detector & Band Center (GHz) & FWHM (arcmin) & $w_E ^{-1}$  [$10^{-6} \mu K^2$]    \\
\hline
Planck & 30 & 33 & 2683 \\
 & 44 & 24 & 2753 \\
 & 70 & 14 & 2764 \\
 & 100 & 10 & 504 \\
 & 143 & 7.1  & 279 \\
 & 217 & 5.0 & 754 \\
 & 353 & 5.0 &6975 \\
\hline
SPIDER & 100 & 58 & 84.4\\
 & 145 & 40  & 47.4  \\
 & 225 & 26 & 395 \\
 & 275 & 21 &1170\\
\hline
CMBPol (EPIC-2m) & 30 & 26 &31.21\\
 & 45 & 17 & 5.79 \\
 & 70 & 11 & 1.48 \\
 & 100 & 8 & 0.89 \\
 & 150 & 5 & 0.83 \\
 & 220 & 3.5 & 1.95 \\
 & 340 & 2.3 &39.46 \\
\hline

\end{tabular}
\label{det-table}
\end{table}

\begin{figure}
\begin{center}$
\begin{array}{cc}
\includegraphics[width=0.5\textwidth]{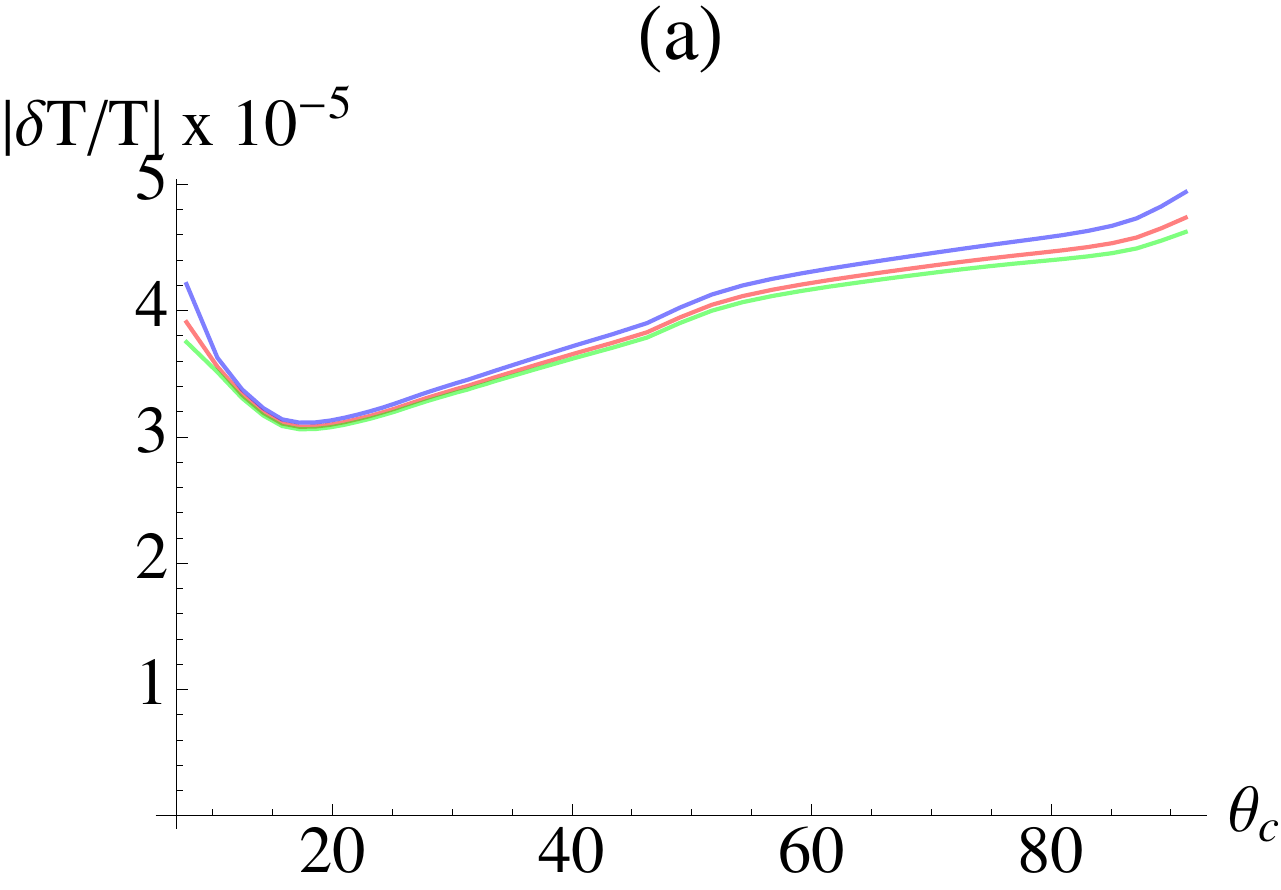}
\includegraphics[width=0.5\textwidth]{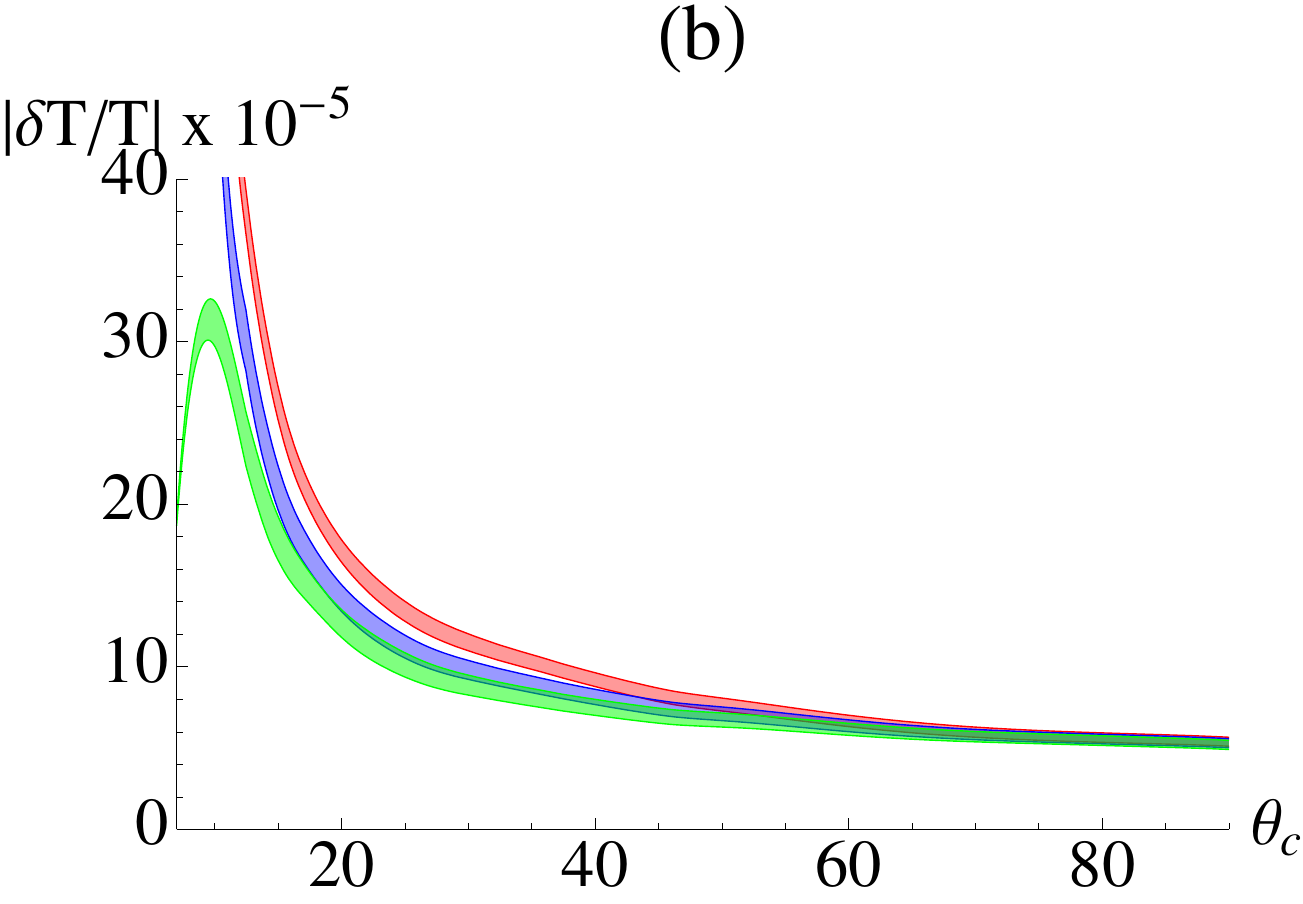}
\end{array}$
\end{center}
\caption{\label{fig-det} $3 \sigma$ detectability thresholds for (a) temperature and (b) E-mode polarization. In each, Planck is red, SPIDER is blue and CMBPol is green. The thickness of the lines in (b) corresponds to using information from the cross-correlation to improve the measurement.}
\end{figure}

For each experiment we do a combined analysis across all bands. Table \ref{det-table} summarizes these numbers (we have assumed the sensitivity to Stokes $I$ is $\sqrt{2}$ that of $Q$ and $U$). For each spot radius $\theta_c$ and experiment we find the minimum $|\delta T / T|$ at the center of the spot necessary such that a detection $A=1$ is $3 \sigma$ away from zero. This represents a conservative estimate of the detectability threshold in temperature and E-mode polarization. 
 We summarize our findings for temperature and polarization in \figref{fig-det}. For polarization we can also choose to use information from the cross-correlation, which increases the sensitivity by allowing us to eliminate the correlated part of the contribution from Gaussian fluctuations. We compute a range of sensitivities using as much information from the cross-correlation as possible to none which is displayed in \figref{fig-det}(b) by the filled region for each experiment.

We can make some general observations. For small spots ($\theta_c \lesssim 10^\circ$), a spot with $|\delta T / T | > 4 \times 10^{-5}$ is likely observable in temperature, but would have to be brighter than $\sim\!2 \times 10^{-4}$ to be observable in polarization. A preliminary search for bubble collisions in the temperature map has been recently carried out \cite{Feeney:2010jj,Feeney:2010dd} using the WMAP-7 dataset. While the analysis detected several anomalous features, it rules out collisions with $| \delta T/ T | > 10^{-4}$. Thus, while the potential still exists to observe small spots in the temperature map, correlating them with polarization signals is likely outside the capabilities of current or near future experiments, at least at the $3\sigma$ level. If we lower our threshold for correlation it may still be possible to obtain some evidence for a collision in the polarization map, though not likely significant enough to exclude other potential explanations.

The situation is considerably more promising for larger spots. With  $\theta_c \simgeq 25^\circ$  much fainter spots are detectable in polarization, and very large spots could be detected almost as easily in polarization as temperature.
The reason for this is clear from \figref{fig-polcross}. Larger, fainter spots have stronger $Q$ (or $E$) -mode polarization over a larger angular range and thus more easily can be seen over the Gaussian fluctuations and detector noise. In addition,  spots larger than approximately $12^{\circ}$ have the distinctive double peak feature on smaller scales. 

This angular size dependence is particularly important because the size distribution of collision spots --- $dN(\theta_c) \propto \sin \theta_c d\theta_c$ --- is a robust prediction of the theory \cite{bubmeas}.  The analysis of \cite{Feeney:2010jj,Feeney:2010dd} was restricted to spots with $\theta_c \simleq 11^\circ$, which, according to this distribution, are much less common than larger spots.  

\section{Conclusions}

We have solved numerically for the real-space evolution of the cosmic wakes produced by collisions between bubble universes, and determined the distinctive  temperature and polarization patterns these cosmic wakes imprint upon the CMB.
These  patterns have a circular symmetry that reflects the near-planar symmetry of cosmic wakes, and the polarization pattern can have a distinctive double-peak structure arising from propagation of acoustic waves in the primordial plasma. We have estimated the detectability of cosmic wakes for several current and future CMB experiments in both temperature and polarization.  Holding fixed the temperature anomaly at the center, increasing the radius of the spot makes it \emph{easier} to detect with polarization, but harder using temperature.
The detection of a cosmic wake would show our observable Universe is one part of a larger multiverse, support the idea of the string theory landscape, and  constitute a groundbreaking discovery of the  nature of the big bang.

\section*{Acknowledgements}
We thank Guido D'Amico, Spencer Chang, Ben Freivogel, Roberto Gobbetti, Gary Hinshaw, Lam Hui, Eugene Lim, Adam Moss, Jonathan Roberts,  Ignacy Sawicki, Roman Scoccimarro, I-Sheng Yang, Matias Zaldarriaga, and James Zibin for discussions.  MK and TSL especially thank UC Davis and the organizers of the workshop ``Bubbles in the Sky" for warm hospitality and a stimulating meeting.   KS thanks the Aspen Center for Physics and Perimeter 
Institute for Theoretical Physics, where parts of this work were completed, for hospitality.
The work of MK is supported by NSF CAREER grant PHY-0645435. The work of KS is supported in part by a NSERC of Canada Discovery grant.  The work of TSL is supported in part by Natural Sciences and Engineering Research Council of Canada and the Institute of Particle Physics.  

\bibliographystyle{klebphys2}

\bibliography{bubble}

\end{document}